# Broadband absorption enhancement in ultra-thin crystalline Si solar cells by incorporating metallic and dielectric nanostructures in the back reflector

**S. Jain, V. Depauw, V. D. Miljkovic, A. Dmitriev, C Trompoukis, I. Gordon, P. van Dorpe, O. El Daif***

**IMEC, PV, Leuven, Belgium**

Corresponding author: ounsi.eldaif@imec.be

## Abstract

We propose a back-reflecting scheme in order to enhance the maximum achievable current in one micron thick crystalline silicon solar cells. We perform 3-dimensional numerical investigations of the scattering properties of metallic nanostructures located at the back side, and optimize them for enhancing absorption in the silicon layer. We validate our numerical results experimentally and also compare the absorption enhancement in the solar cell structure, both with quasi-periodic and random metallic nanostructures. We have looked at the interplay between the metallic nanostructures and an integrated back-reflector. We show that the combination of metallic nanoparticles and a metallic reflector results in significant parasitic absorption. We compared this to another implementation based on titanium dioxide nanoparticles which act as a lambertian reflector of light. Our simulation and experimental results show that this proposed configuration results in reduced absorption losses and in broadband enhancement of absorption for ultra-thin solar cells, paving the way to an optimal back reflector for thin film photovoltaics.

## (1) Introduction

Ultra-thin film (<5μm) and thin film (<100μm) crystalline silicon solar cells are promising solutions for the issue of high production costs of solar cells. Ultra-thin film (<5μm) cells, besides the cost-reduction advantage, offer various advantages like faster production, higher open circuit voltage[1], and enhanced charge collection efficiency[2]

However, with reduced thickness of the crystalline silicon (c-Si) layer, absorption decreases considerably due to the low absorption coefficient of c-Si (due to its indirect band gap); it requires ~1mm of cSi to completely absorb light of 1100nm wavelength and this results in low efficiency of thin film solar cells. Thus as we go towards thinner substrates, the importance of light trapping increases significantly. Currently, to address the problem of low light absorption in standard industrial solar cell structures (~ thickness 180 μm) solutions like Anti Reflection Coatings (ARC) and surface texturing at micron scale are used. Typical surface texturing has a feature size of 10-15 μm and around the same thickness of the material[3] is lost, making it unsuitable for application in thin film and ultra-thin film solar cells. Some groups have proposed submicron surface texturing[4],[5],[6], however this results in very high surface roughness and hence it enhances losses due to surface recombination. As for ARCs, they are not sufficient for increasing absorption in thin film cells, for they only enhance transmission by increasing the angle of light entering the substrate. But as the semiconductor layer is thin, light is not fully absorbed in one pass, and it needs to be redirected back to the semiconductor in an efficient manner.

Ever since the path breaking work of Michael Faraday [7] on the problem of colour of colloidal gold solution, there has been growing interest in optical properties and behaviour of metallic structures at nanoscale. Noble metal nanostructures are known to support localized surface plasmon's[8], i.e. electromagnetic modes arising at the interface between a metal and a dielectric. This plasmonic behaviour arises due to the collective oscillation of confined conduction electrons with the incident electromagnetic wave. For structures having dimensions smaller than the wavelength of light impinging on them, 20 the conduction electrons start oscillating in phase, resulting in charge polarisation at the surface. This charge polarisation
21 builds up a restoring force which sustains the resonance at specific wavelengths, giving rise to a resonantly enhanced field
22 throughout the particle and also outside of it. This results in large scattering and absorption cross section of plasmonic
23 metallic nanoparticles and enhances the electric field in the surroundings. The plasmonic behaviour, damping and
24 strength of the plasmonic excitation, is a function of the shape, size, and material of the particle, of the dielectric function
25 of the surrounding medium and nearby material[9] and can thus be tuned to the desired spectral range. Incorporation of
26 plasmonic metallic nanostructures has the potential to overcome the problem of light trapping in ultra-thin film solar cells
27 by reducing reflection or by redirecting light towards silicon in efficient manner and is a widely studied field[10], [11]. Two
28 main approaches applied for using the plasmonic particles depending on the cell architecture and material for solar cells
29 are: 1) incorporating the plasmonic nanoparticles directly into the semiconductor layer, an approach which takes
30 advantage of the near field enhancement[12] and [2]) using the plasmonic nanoparticle on the front or the back side of the
31 solar cells, an approach which employs the resonant scattering property of plasmons along with their directional
32 scattering property[13].

33 Earlier studies suggested that using plasmonic nanostructures at the front side of the solar cell results in a decrease in its
34 energy-conversion efficiency [14, 15] . The reason attributed to this is that, for frequencies greater than the plasmon
35 resonance of the metal nanoparticles, the destructive interference between transmitted and scattered electromagnetic
36 wave results in Fano resonance, causing an increase in reflection[16, 17] and hence reducing the external quantum efficiency.
37 To avoid this loss, the trend has shifted from using the plasmonic nanostructure at the front to using them at the back,
38 which uses preferential scattering properties. Preferential scattering phenomena arise when the near-field around the
39 dipole interacts with the high-index surrounding. The closer the particle is to the high-index material, the larger the
40 asymmetry is in the angular distribution of scattered light, thus enabling us to use the particles at the back side
41 configuration and scatter light towards the semiconducting layer.

42

43 In this study, we address both the problem of low light absorption in ultrathin substrates and of parasitic/Ohmic
44 absorption in the flat metal back reflector. First, we optimize the back reflector architecture by employing metallic
45 nanostructures that exhibit plasmonic characteristics, so as to scatter light back into silicon. We then replace the flat
46 metallic back reflector (i.e. flat aluminum layer) by a dielectric back surface reflector for diffusively reflecting light back
47 towards silicon, thus having the advantage of randomizing the direction of reflected light and also avoiding any parasitic
48 absorption as in the case of the flat metallic back reflector.

49 ## (2) Methods and Techniques

50 ## (2. A) Structures chosen

51 We use one-micron-thick epifree crystalline silicon (c-Si) layers on glass, developed in house[18]. This technology gives the
52 possibility of fabricating solar cells on exceptionally thin monocrystalline sililcon films[19]. The silicon layer is bonded to a
53 transparent superstrate after deposition of an antireflection coatingand contacting is therefore completely done through
54 the rear.

55 In order to isolate the effect of the back reflector, we first studied the effect of the plasmonic nanostructures on a minimal
56 stack of silicon on glass with intermediate ARC. Once we studied its effect and optimized it for absorption enhancement,
57 we extrapolated and adapted the optimal back reflecting structure to an active solar cell stack, with a 40nm SiNx
58 antireflection coating and a back-side heterojunction emitter stack (amorphousSi and ITO [19]).

59 In each case, we use as a reference the planar structure (either silicon alone or the cell stack) with an aluminum metal
60 reflector directly deposited on the back, as shown in Figure 1.

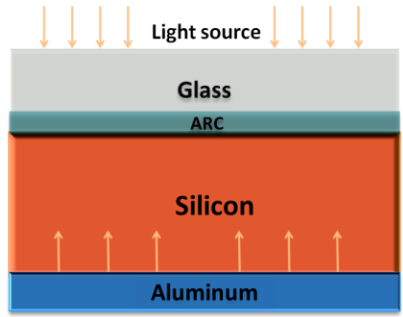

*Figure 1: Reference structure with ARC and aluminium layer as back reflector*

The scattering cross section of the plasmonic nanostructure is maximum at its plasmonic resonance frequency. But when an array of particles is considered, the term scattering cross section for a plasmonic structure is not valid, as the interactions of the particles' plasmonic modes with nearby particles change the overall behavior and broaden the plasmon resonance region. As our objective is to enhance absorption in silicon throughout the whole spectrum, we do not study specifically the resonance region.

Because of their low surface coverage, the plasmonic nanostructures do not reflect all of the light reaching the back side of the stack and therefore need to be assisted by a back full surface reflector. Therefore, in addition to the plasmonic structures we consider a full aluminum layer or a coating of TiO$_2$ nanoparticles to be used as back surface reflectors. Both structures are described below.

## (2A-1) Back reflector scheme with plasmonic nanoparticles and full metal layer

We use plasmonic metallic nanostructures to scatter the light, which is not absorbed in one pass, back towards the silicon, thanks to their preferential scattering characteristic. Silver was chosen as the material for plasmonic nanostructures based on the fact that it is less absorbing and has the plasmonic resonance in the desired wavelength range [20], [21]. We choose silver nanodisks as a plasmonic nanostructure due to their higher scattering cross section [17]. The plasmonic back reflector scheme consists of a SiO$_2$ spacer layer adjacent to silicon, followed by a random or quasi periodic array of silver nanodisks, a SiO$_2$ capping layer and an aluminum coating as a back surface reflector (shown in Figure 2(a)).

## (2A-2) Back reflector scheme with plasmonic and dielectric nanoparticles

A dielectric material with a high refractive index strongly reflects light, so we choose a titanium dioxide (refractive index ~2.73) nanoparticle coating as an alternative to an aluminum back surface reflector, taking advantage of both its good scattering and reflective properties [22]. Since simulating a back reflector scheme with densely packed dielectric nanoparticles of different sizes is not practical, we compare experimentally our concept of dielectric nanoparticle coating as shown in Figure 2(b) with an aluminum layer, to be used as a back surface reflector for the back reflector scheme with plasmonic nanostructures.

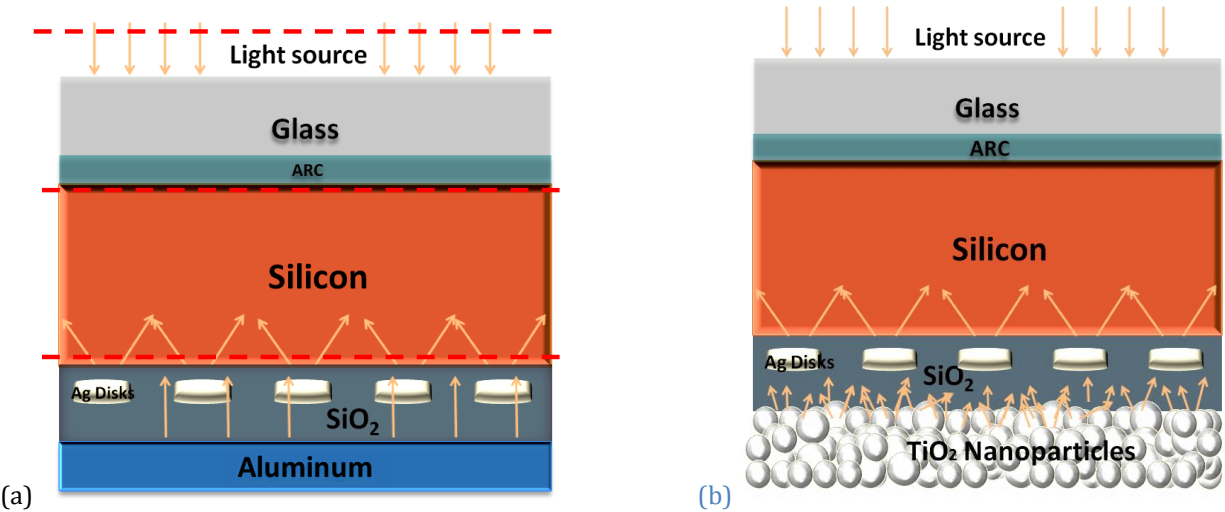

8    (a)                                                      (b)

9    *Figure 2: (a) Optical stack with plasmonic back reflector scheme in presence of (a) aluminium layer (b) dielectric*
10   *nanoparticle coating, as back surface reflector. The red dashed lines in (a) shows the placement of monitors(in 3D*
11   *simulations)  for measuring absorption in different layers.*

## (2 B) Fabrication & Characterisation techniques

### (2B-1) Fabrication of the 1µm-c-Si on glass with intermediate ARC

The 1-µm-thin c-Si films coated with a thin layer of SiN$_x$ are fabricated by a layer-transfer process and are bonded on a glass carrier[20]. The c-Si film is first formed by the *empty-space-in-silicon-technique*[23] which consists of annealing above 1000 °C a wafer patterned with an array of cylindrical macropores that close, and eventually all merge to form an overlaying film. In the present work, the thin film was formed by annealing at 1130 °C under 1 atm of H$_2$. After annealing, the suspended film was coated by a 40-nm layer of SiNx formed by PE-CVD. It was then anodically bonded to a glass substrate by applying 1000 V at 320 °C. Finally, the film was detached from its parent wafer by pulling softly.

As we had to adapt the fabrication flow [19] in order to be able to process the rear side, the front side ARC had to be set between the glass substrate and the thin silicon layer, leading to difficulties in controlling its thickness, which is influenced by the bonding step. This thickness inaccuracy impacts the short wavelength part of the various measured spectra. However, as the part of the spectrum of interest for the back side reflector is here at wavelengths > 500nm, the ARC non-homogeneity does not interfere with the phenomena we are studying.

### (2B-2) Fabrication of quasi periodic silver nanodisks

The quasi periodic arrays of silver nanodisks have been fabricated by hole-mask colloidal lithography (HCL)[24]. A typical nanodisks arrangement exhibits short-range order, but no long-range order, as can be seen in Figure 3(a).

### (2B-3) Fabrication of random arrays of silver nanoparticles

Random arrays of silver nanoparticles were fabricated based on the coalescence of a thin metal film. A Ag film of thickness ~14nm was deposited on an optical stack with the SiO$_2$ spacer layer already deposited and then annealed for 90 minutes under N$_2$ atmosphere at 220⁰ C. Annealing under the above-mentioned conditions results out of thermal stress in the formation of a random array of silver nanoparticles, as shown in the Scanning Electron Microscopy (SEM) image in Figure 3(b). The size distribution is broader than for the HCL technique[25].

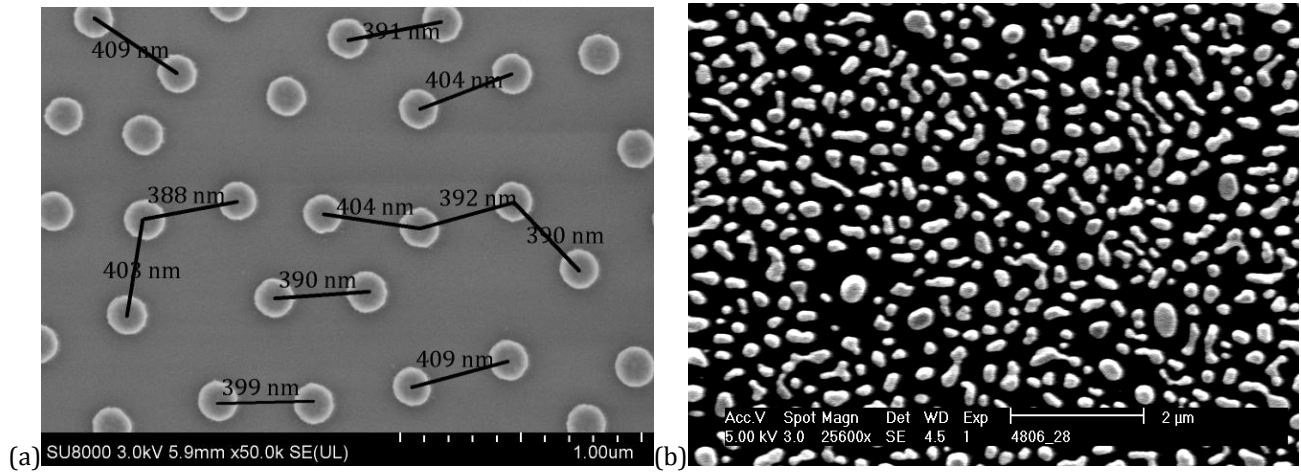

*Figure 3: SEM image of [a] the quasi periodic array of nanodisks fabricated by HCL and [b] the random array of Ag nanoparticles formed by thermal annealing of a Ag film on the optical stack.*

### (2B-4) Fabrication of the dielectric nanoparticles coating

The dielectric nanoparticles coating was fabricated by dispersing $TiO_2$ nanoparticles (combination of particles sizes of 405nm, 320nm and 220nm) in propylene glycol methyl ether acetate. After dispersion, the solution was dip coated on the sample and annealed for 20min at $60^0C$. The coating on the glass slide was used for comparing the reflection with the aluminum layer (also deposited on glass). For using in conjugation with the plasmonic nanoparticles, this coating was mechanically stacked on the backside of the structure.

### (2B-5) Characterisation

We first characterized the particles by SEM in order to check their size and spatial distribution. We then optically characterized the samples, allowing the deduction of the reflectance spectra of the various back reflectors studied, as well as the various stacks. We measured transmittance as well when relevant. These two measurements allow us to deduce the absorption spectra. We used there an integrating sphere, which sums the reflection signal in the whole half space in front of the sample. The light incident on the sample was scanned from 400 to 1200nm of wavelength, corresponding to the limits given both by the solar spectrum and silicon's absorption range.

### (2 C) Simulation setup

We chose the Finite difference time domain[26] (FDTD) method and used a commercially available package: Lumerical FDTD solution[27], for simulating the optical behavior of our solar cell stack. Three Dimensional (3-D) simulations were performed by using periodic boundary conditions in x and y dimensions to replicate the actual sample. Perfectly matched layer (PML) boundary conditions were used in the direction of propagation of light i.e. in z dimension, in order to account for the case that the light which is not absorbed by the structure is not reflected back in the simulation region. For high accuracy and detailed insight on the effect of parameters on the absorption throughout the chosen spectrum, the simulations were performed from 300nm to 1200nm with a wavelength step of 0.9nm. The refractive index values for silicon, silver, silicon di-oxide and aluminum were taken directly from Lumerical database, whereas the refractive index of silicon nitride was measured using ellipsometry.

For calculating the absorption in any specific layer, the monitors were placed at the start and end of the layer (as shown by red dashed lines in Figure 2(a)). By calculating the difference in the amount of light entering and exiting the layer, we measured the absorption in that specific layer. Since a metal back surface reflector is present after the plasmonic back reflecting scheme, no light is crossing the simulation region in the +z dimension, and all of the light that is not absorbed is reflected back. The absorption in the whole structure is calculated by subtracting the data collected from the motor placed before the source (as shown in Figure 2), from the amount of light falling into the substrate. This arrangement of monitor placement allows us to measure the absorption in the structure qualitatively and thus can be used to calculate it in any given layer.

The scattering behavior of the plasmonic nanostructure is a function of the distance of the particle from the high index medium (spacer layer), the dielectric function of the surrounding and nearby medium (material in which the particle is embedded), as well as the shape and size (height and diameter) of the nanoparticle itself. The resonance wavelength is also affected by interparticle spacing between nanodisks, which in the case of HCL samples is defined by electrostatic repulsion between charged colloidal beads [24]. Looking at the characteristic interparticle distances from Figure 3(a), and from the fact that using a period below 400nm results in very insignificant light trapping (as no light couples with silicon's higher order diffraction modes[28]), we choose 400nm to be the interparticle spacing and optimized the other parameters. All of these parameters (shown in Figure 4) were optimized for maximizing absorption by the silicon layer.

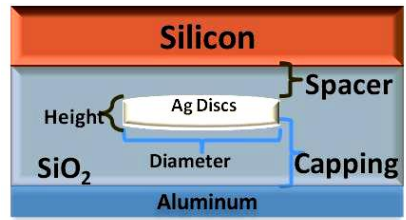

*Figure 4: Parameters of the back reflector scheme optimised in this study.*

The enhancement in absorption in silicon does not translate equally in an improved $J_{sc}$ (short circuit current), as the incident solar spectrum is not flat and the energy of photons decreases with an increase in wavelength. To have qualitative insights on the effect of incorporation of the plasmonic back reflector scheme in the thin film cells, absorption is integrated over the AM1.5 solar spectrum so as to calculate the $J_{sc}$ of the device using the equation below

$$ \tag{1} $$

where q is the elementary charge, $\lambda$ is the wavelength, $h$ is the Planck constant, c is the speed of light, QE is the quantum efficiency (collection efficiency in our case, as we are calculating absorption in the silicon layer only), S is the weighted sun spectrum (AM1.5 spectral irradiance) and A is the absorption in the material. Considering the collection efficiency in the silicon layer as unity i.e. assuming that every absorbed photon produces one charge-carrier pair, the maximum achievable current $J_{scmax}$ is calculated and is used as figure of merit for optimizing the parameters for the plasmonic back reflector.

## (2C-1) Validation of simulations

Minor inconsistencies between simulation and experiments can arise both in the trend and absolute value of the parameter under investigation due to various reasons. One of these is the fact that simulations consider a coherent light source whereas the real coherence length is limited to a few microns. Secondly, the refractive indices of the materials were taken from Lumerical's data base, which can sometimes slightly differ from real values. Thirdly, the roughness of the surfaces and interfaces determines their reflection; in simulation we choose perfect and flat surfaces, which is of course not the case experimentally. Finally, experiments always carry a certain degree of uncertainty in the thickness of the deposited material. Considering all these factors, before utilizing the simulations for studying and optimizing the parameters for back reflector schemes, we compare the results of simulations with the experiments to ascertain the degree of inconsistency. We choose two structures for the validation of simulations. The first structure consists only of a 1μm thin c-Si layer on glass while the second structure is a 1μm thin c-Si layer on glass with an aluminum back reflector. Despite the various factors that can result in any difference between the simulation and experimental data, our simulation results are within acceptable limits, as can be seen from the comparison in Figure 5(a) and 5(b). We attribute the slight discrepancies, mainly noticeable at short wavelength, to the error on the refractive index data which was input into the model, but. as this wavelength range is not our target, this will not impact our results.

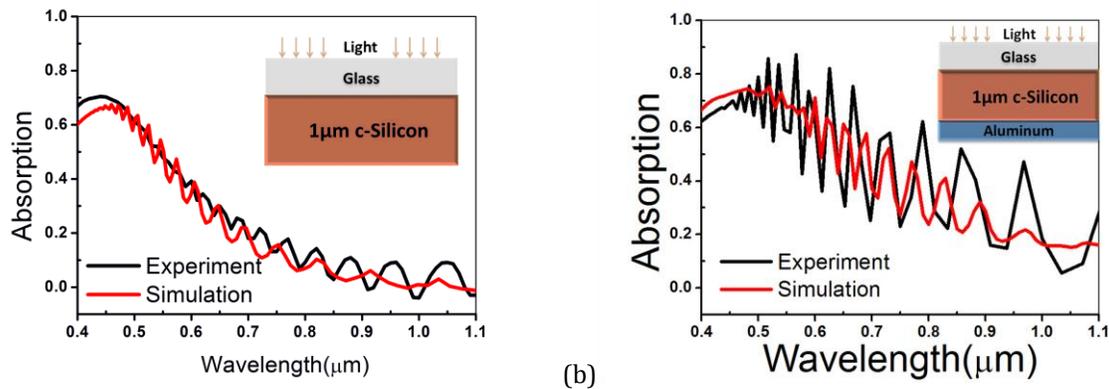

(a)  (b)

107 *Figure 5: Comparison of experimental and simulation results for the absorption in the structure consisting of (a) Silicon on*
108 *glass (b) Silicon on glass with aluminium as back surface reflector .*

### 109  (3) Simulation results and discussion

110 We scan hereafter the various relevant parameters of the plasmonic back reflector and extract a set of optimised
111 parameters.

### 112  (3 A) Optimization of plasmonic back reflector design with metal layer at the back

113 The dimensions of the nanodisks (height and diameter), the distance of plasmonic nanodisks from silicon (spacer layer),
114 and the distance between the aluminum and the nanodisks (capping layer) (Figure 4)) were scanned individually within
115 300-1200nm wavelength range for maximizing the figure of merit, i.e. the maximum achievable current. The absorption in
116 the silicon layer was extracted for various values of parameters and $J_{scmax}$ was calculated and compared. Spacer thickness
117 of 13nm, diameter of 205nm, height of 55nm and capping layer 110nm thick were chosen as starting parameters based on
118 extrapolation of previous studies performed for the nanodisks situated at the front [16, 29], and earlier calculations carried
119 out for nanodisks situated at the back [30]. A first rough scan was performed, followed by a finer one, and results are
120 presented here below.

### 121  (3A-1) Optimization of the capping layer:

122 The capping layer thickness influences the Fabry–Pérot (FP) cavity resonances, which are standing waves between the
123 silicon surface and the metal back reflector. Depending on the (spectral and spatial) positions of the minima and maxima
124 of the field it can enhance the absorption in the particles and the aluminium film or enhance the scattering efficiency of
125 the metallic nanodisks. The ideal thickness cannot be chosen by considering only the FP modes, because the presence of
126 particles strongly enhances the electric field in the region and thus changes the overall behaviour, which is why we do not
127 split the thickness study of the back dielectric layer from the NPs study. The thickness of the spacer was kept constant at
128 13nm while the capping layer thickness was scanned form 100nm to 560nm, with a step of 20nm. The upper limit of
129 560nm was chosen in order to ensure the scanning of at least a free spectral range between the ground mode and the next
130 resonance. A fine tuning was done near the values of thicknesses which gave maximum $J_{scmax}$ enhancement. From the $J_{scmax}$
131 curve shown in Figure 6(a), 210nm of capping layer scatters the light into silicon most effectively and gives a $J_{scmax}$ value
132 of **24.76mA/cm²**.

### 133  (3A-2) Height of the nanodiscs

134 When the disks are too thin there is a high parasitic absorption and when they are too thick, the plasmonic resonance
135 wavelength for dipole oscillations red shifts and higher order modes appear at lower wavelengths [31]. The thickness of the
136 nanodisks was scanned from 20nm to 70nm, for every 10nm, while keeping the thickness of the capping layer constant at
137 210nm. We observe that by decreasing the height of nanodisks below 50nm, it decreases the scattering efficiency.

138  Increasing the thickness of the nanodisks results in blue shifting of their surface plasmon resonance (SPR) frequency. A
139  50nm height of nanodisks scatters the light into silicon most effectively and gives a $J_{scmax}$ value of **25.01mA/cm²**.

### (3A-3) Diameter of the nanodiscs:

141  The diameter of silver nanodisks was scanned from 100nm to 300nm for every 20nm, keeping their thickness constant at
142  50nm, and the thickness of the capping the layer at 210nm. Decreasing the diameter below 100nm increases the Ohmic
143  absorption rapidly as the particle size becomes too small with respect to the wavelength of incident light. For the particles
144  with bigger diameter (~300nm), the resonance shifts away from the desired part of the spectrum[31]. A 200nm diameter of
145  nanodisks scatters the light into silicon most effectively and gives a $J_{scmax}$ value of **25.27mA/cm²**.

### (3A-4) Optimization of spacer layer:

147  With a thin spacer layer, the effect of silicon on the scattering properties of the plasmonic nanostructures will be
148  maximum as, the closer the structure is to the high index medium, the higher is the effect on the homogeneity of its
149  surrounding and hence on the preferential scattering properties.  The effect of the spacers thickness was studied on the
150  structure from 0nm (i.e. nanodisks directly on the silicon) to 60nm, for every 5nm. The upper limit of 60nm was chosen
151  keeping in mind that after a few tens of nanometers the effect of a nearby high index material on the scattering properties
152  of the nanostructures decreases significantly, thus having negligible effect in the scattering pattern.  A second scanning of
153  thickness was done near the values which gave the highest $J_{scmax}$ values. A 10 nm of spacer layer results in the highest
154  enhancement in the silicon layer with $J_{scmax}$ value of **25.43mA/cm²**. After the spacer thickness of ~20nm, silicon's ability
155  to modify the surrounding environment around the plasmonic particle starts decreasing and keeps on decreasing as the
156  plasmonic nanostructure moves away from silicon. Another reason is the evanescent nature of the surface Plasmon field,
157  which decreases very rapidly from the surface of the nanoparticle. Thus the further the nanoparticle is from silicon, the
158  weaker is the coupling of the enhanced field into the silicon. This can be seen from the decrease in the $J_{scmax}$ values for the
159  values higher than 20nm in Figure 6(d).[32]

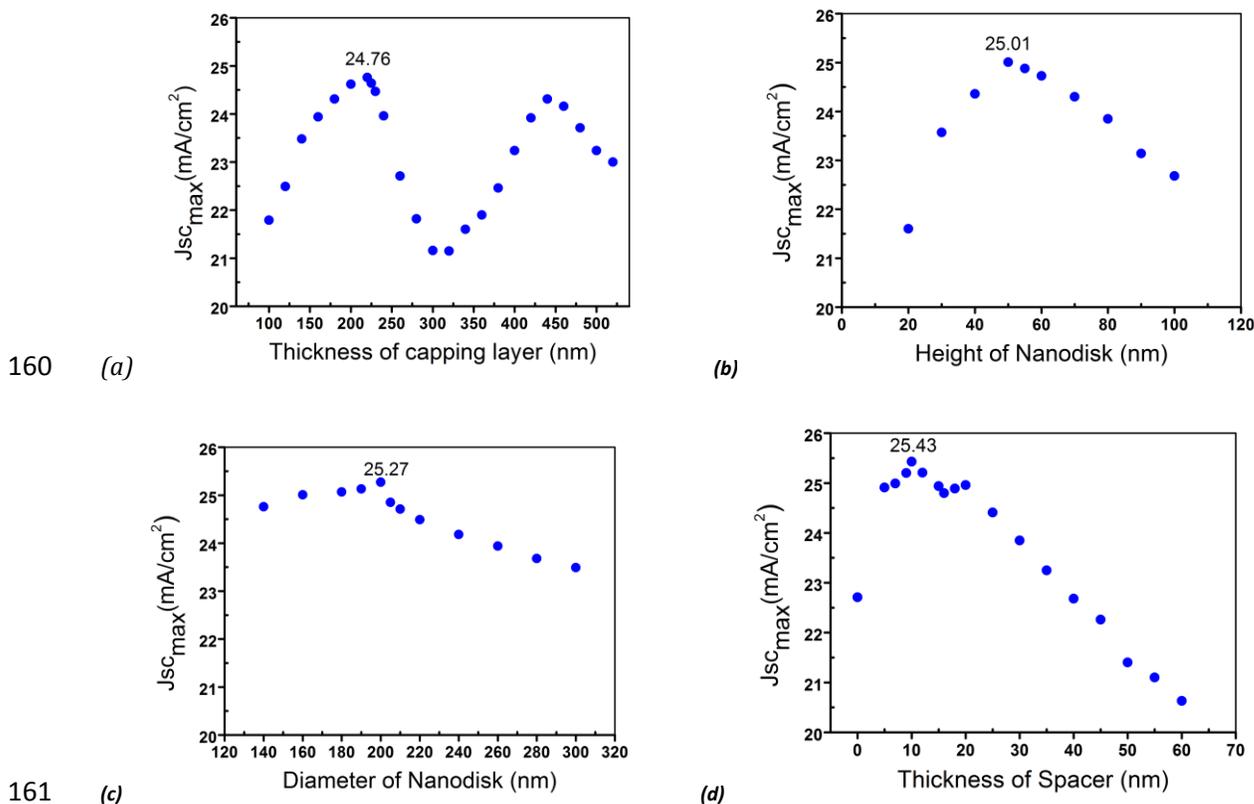

160  *(a)*
161  *(c)*     *(d)*     *(b)*

162  *Figure 6: $J_{scmax}$ values in function of various parameters (a) Thickness of capping layer (b) Height of Nanodisk (c) Diameter of*
163  *nanodisks (d) Thickness of spacer*

### 164  (3A-5) Summary of the parameter study and discussion

165  We were able to achieve significant broadband enhancement, illustrated in Figure 7(a) that shows the comparison of
166  absorption in silicon between the structure with the optimized back reflector scheme and the reference structure.

167  The $J_{sc\,max}$ is increased from **13.6 mA/cm²** for the reference structure without ARC, and from **15.86 mA/cm²** for the
168  reference structure with ARC, to **25.43 mA/cm²** for the structure with the optimised back reflector scheme and an ARC,
169  which amounts to **~60.3%** AM1.5 photon absorption enhancement in silicon. We also compare the absorption
170  enhancement in the silicon layer by comparing the number of photons absorbed from the real AM1.5 solar spectrum
171  (Figure 7b); this gives both a qualitative and a quantitative aspect of absorption enhancement.

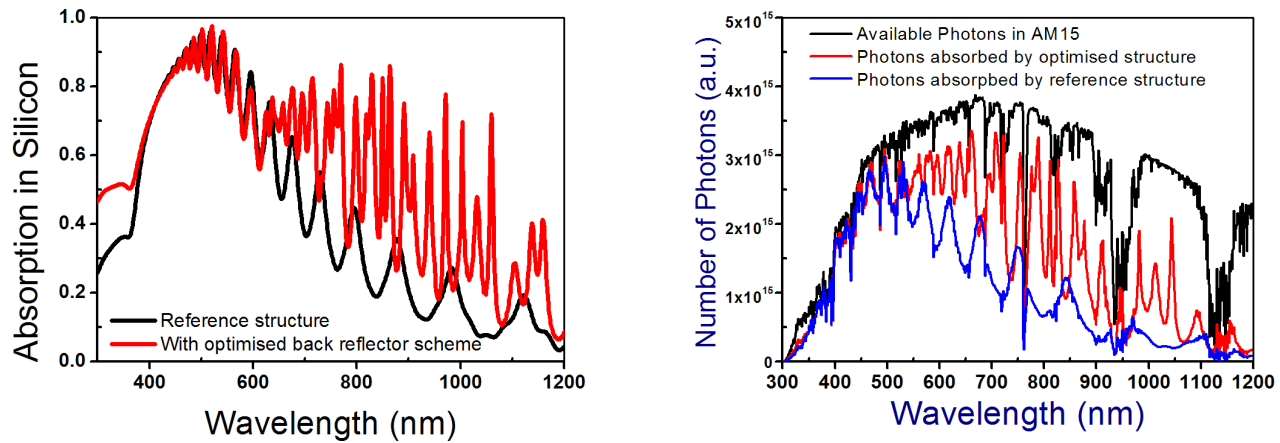

172

173  *Figure 7:(a) Integrated absorption in silicon with optimized back reflector (black) and reference structure (red). (b) Number*
174  *of photons absorbed in the silicon layer in the reference structure (red) and in the optimised plasmonic solar cell structure*
175  *(blue). The black curve shows the number of photons present at AM1.5 conditions.*

176  By comparing the red curve (optical stack with optimized plasmonic back reflector scheme) with the blue curve
177  (reference solar cell structure), we clearly see that there is broadband enhancement in absorption throughout the
178  spectrum, especially from ~550nm to 1100nm.

### 179  (3 B) Absorption in metal back reflector

180  There is always a significant parasitic/Ohmic absorption (light which is absorbed by the structure and does not result in
181  charge carrier generation), both in the plasmonic nanostructures and the metal layer which is used as back surface
182  reflector. We cannot avoid the parasitic absorption in our plasmonic nanodisks beyond a certain limit as we have already
183  optimized the parameters to maximize the useful absorption. But we can still minimize parasitic absorption in the
184  aluminum layer that is used as flat back reflector layer.  We can quantify this absorption in the aluminum layer in
185  combination with the plasmonic nanostructures. The parasitic absorption in the presence and absence of the flat
186  aluminum back reflecting layer is compared in Figure 8(a).  We clearly see a high amount of absorption in the system
187  combining nanodisks and aluminum layer as back surface reflector. This suggests that the presence of the aluminum layer
188  may enhance the parasitic absorption in the plasmonic nanoparticles, via coupling of the exponentially decaying field of
189  the metal film with the Plasmon modes of the nanoparticles embedded in dielectric[33]. To have further insight, we compare
190  the absorption in the nanodisks in presence and absence of a flat aluminum layer, as shown in Figure 8(b).  We also
191  quantify the absorption in the aluminum layer alone as shown in Figure 8(c).

192  The comparisons clearly show that there is almost no enhancement of the parasitic absorption in the plasmonic silver
193  nanodisks by adding the flat aluminum layer. The major fraction of the light in the wavelength range of 550-850nm is lost
194  due to the parasitic absorption in the flat aluminum layer. Essentially, what happens is that the nanoparticles scatter light
195  to the region with the largest photonic density of states (or the largest index). Silicon has a large index, but the surface
196  plasmons on the Al surface also have a large effective index. The closer we put the Al to the particles, the more effective

197 the coupling between the particles and the Al layer. Which explains that the extra absorption will take place in the Al, not
198 in the Ag particle. If we can avoid this loss and redirect this light into the silicon, we will considerably enhance absorption
199 in silicon.

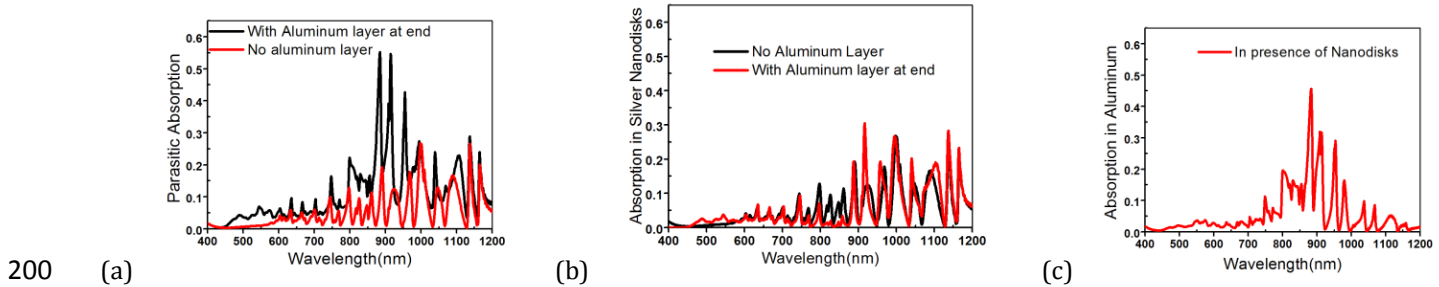

200 (a)           (b)           (c)

201 *Figure 8: (a) Parasitic absorption in the presence (black) and absence (red) of aluminium layer but always with*
202 *nanoparticles on the back, (b) absorption in the silver nanodisks in presence (black) and the absence (red) of aluminum layer,*
203 *(c) absorption in the aluminum mirror in presence of silver nanodisks.*

### (3B-1) Replacing aluminium with perfect metal boundary

205 We now simulate the structure with an optimized back reflector scheme and replace the aluminum layer by a perfect
206 metal (using metal as boundary condition in +z direction). This metal boundary condition means 100% reflection and no
207 parasitic absorption. This allows giving an idea of the possible gain which would be brought by a non-absorbing reflector
208 behind the nanoparticles. Figure 9 compares absorption in silicon with the aluminum layer and with a perfect metal (non-
209 absorbing) layer as back surface reflector.

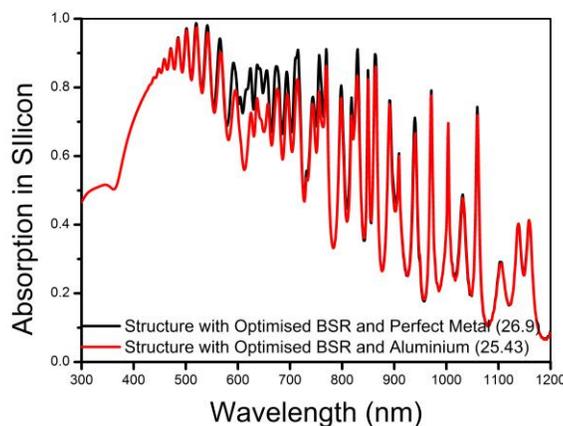

211 *Figure 9: Absorption in silicon with a perfect metal (black) and an aluminium layer (red) on the back.*

212 As there is no absorption in a perfect metal, the light which was previously lost because of these parasitic absorption in
213 aluminium is sent back towards silicon, hence increasing absorption in the silicon layer. One can see that this happens at
214 the plasmonic resonance, which corresponds to the maximum in parasitic absorption in the metal stack (see Fig. 8 a and
215 c) The structure with the optimised back reflector scheme with a perfect reflector gives a $J_{scmax}$ of **26.9mA/cm²**, whereas
216 with an aluminium layer as back surface reflector the $J_{scmax}$ is **25.43mA/cm²**. Thus by using a perfect metal as back
217 surface reflector with plasmonic nanoparticles we were able to further enhance the maximum achievable current by
218 **~69.6%**. In practice, the best way to approach an ideal metal reflector is to use a multilayer dielectric Bragg reflector, as
219 further discussed in the experimental part.

220

221

| Structure | Jscmax (mA/cm$^2$) |
|---|---|
| Reference | 15.9 |
| Reference with ARC | 18.4 |
| Structure with ARC and optimised back surface reflector | 26.9 |

Table: 1 Best currents opbtained

## (4) Experimental results and discussion

### (4-1) Experimental characterisation of the numerical optimal

After validation of our model with the reference optical stack (Section 2C-1), we now validate our simulations for the plasmonic back reflector scheme. We fabricated the plasmonic back reflector scheme on the back side of a 1μm thin c-Si layer with nanodisks of height 50nm, ~ 200nm diameter, ~15nm spacer and capping 100nm. The absorption in the structure was compared with the simulated results, shown in Figure 10.

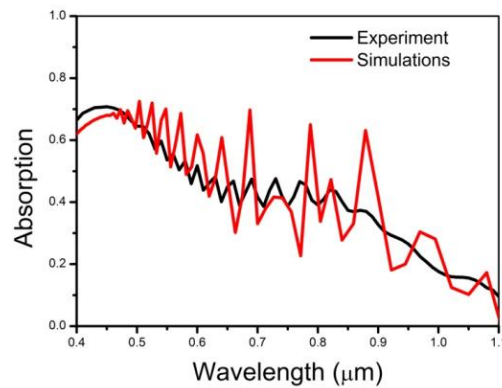

*Figure 10: Comparison of absorption in the structure with the plasmonic back reflector scheme without a metal layer at the back; simulations (red), experiment (black).*

Simulation and experiments are in good agreement. The difference at short wavelengths in the experimental and simulation data can reasonably be attributed to the previously discussed refractive index discrepancy. Another major factor is that in the simulations, the interparticle spacing was fixed to be 400nm, while in the experiment it varies with ±20nm (as seen in Figure 3(a)), this yields sharper resonances in the red part of the simulated spectrum.

246  **(4-2) Beyond the numerical optimal: using dielectric nanoparticles coating as a back reflector**

247  Absorption in the aluminum layer in the back reflector scheme is significant and is highly undesirable as it lies mainly in
248  the wavelength region of interest. In order to overcome this, we have seen previously that a perfect metal would be ideal,
249  although of course not available. However, a flat reflector, ideal or not, can have the drawback of enhancing the coupling
250  between the Fabry-Pérot cavity and the plasmon modes which is the cause of this high absorption in the back reflector, as
251  can be seen from Figure 8(c). Therefore beyond having an ideal reflector, a loss of coherence would help avoiding any
252  interference and therefore any kind of coupling between the light reflected by the back reflector and by the plasmonic
253  behavior of the nanostructures. This can be achieved if the reflection from the back surface reflector is not specular but
254  diffuse. Therefore, we propose to replace the metal layer in the back reflector scheme by a coating of dielectric
255  nanoparticles which will scatter back light thanks to Mie scattering[34], as shown in Figure 2(b). For diffused scattering of
256  light throughout the spectrum, we chose a combination of three different particle sizes of titanium dioxide i.e. 405nm,
257  320nm and 220nm in order to achieve scattering on a broad spectral range.

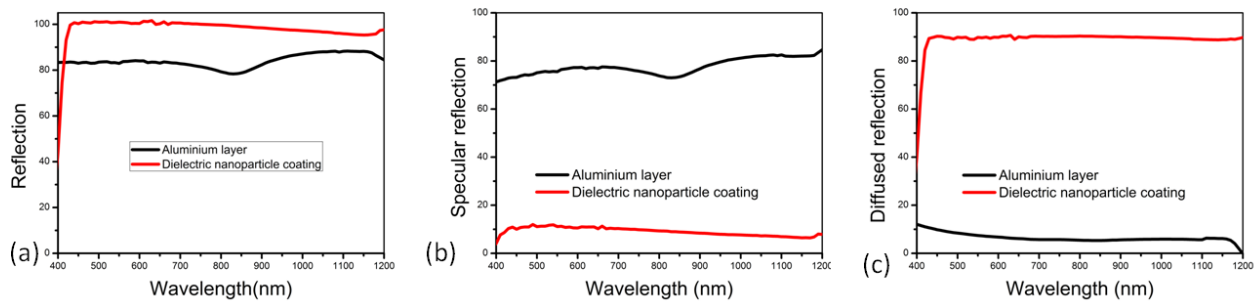

258

259  *Figure 11: Comparison among difference components of reflection from the Aluminium layer (black) and the dielectric*
260  *nanoparticle coating (red) (a) Full reflection, (b) Specular part of reflection, (c) Diffuse part of reflection.*

261  From Fig.11 we see that the dielectric nanoparticles coating scatters ~ 80% of the reflected light compared to ~10 % for a
262  flat metallic back reflector, thanks to the lambertian nature of the reflection profile. Thus it is very effective in
263  randomising the direction of light that is reflected back into the silicon and should enhance absorption as compared to the
264  aluminium layer. The fact that the reflected light is diffuse is expected to avoid any coupling between the plasmon
265  resonance and cavity modes between the silicon and the flat metallic back surface reflector.

266  **(4-3) Experimental comparison of optimized back reflection schemes**

267  We fabricated the cell stack with an optimized back reflector scheme, following the simulations presented, for optical
268  characterization. We chose to fabricate this optical demonstrator with indium tin oxide (ITO) surrounding the metal
269  nanoparticles instead of SiO$_2$,keeping in mind the future need for contact formation in actual devices.

270  The various parameters of the back reflecting stack are aimed to correspond to the results of the numerical optimisation
271  within the experimental uncertainties. We tried two sets of nanoparticles: quasi-periodic and random plasmonic
272  nanoparticles. The particles fabricated by the HCL process have a diameter of ~200nm and an inter-particle distance of ~
273  400nm, while the particles fabricated by an anneal step have a broad size distribution. Figure 12 compares the absorption
274  for the structure without any back reflector (which will from now be used as a reference), the structure with aluminum
275  back-reflector, the structure with quasi-periodic nanoparticles, and finally with random nanoparticles.

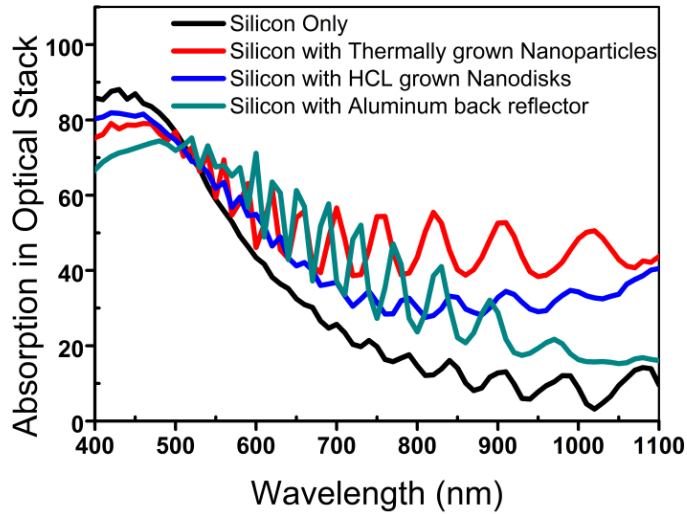

*Figure 12: Absorption in the optical stack with ARC; (black) Silicon only, (blue) Silicon with HCL grown nanodisks only, (red) Silicon with thermally grown nanoparticles only (green) Silisonc with a flat Al back reflector.*

Absorption measurements in the structure without back surface reflector show expected results. The stacks with nanoparticles on the back show an enhanced absorption, thanks both to the scattering and reflective behavior of the nanoparticles. The higher absorption in the stack with fully random (annealed) particles can be attributed both to the higher surface coverage (as can be seen on Figure 3(b)) and to the broader scattering region due to the broader size distribution. The difference in absorption in wavelength below 500nm is due to the difference in thickness of ARC due to experimental errors as explained above, and not related to the back-reflector.

In order to demonstrate the scattering character of the plasmonic particles, and to differentiate it from a simple reflector with a partial surface coverage, we proceed in two ways:

1. We add a metallic planar back reflector on the back of the planar structure without nanoparticles (green curve in Fig. 12) and compare it with the structures with the nanoparticles. One can observe that the absorption of the planar structure is enhanced thanks to a doubling of the optical path length of the red and near-infrared photons, causing a proportional enhancement of the absorption in this spectral range. On the other hand, the structures with nanoparticles show a non-constant increase of absorption showing a scattering resonance acting mainly in the red-near-infrared spectral region.

2. We showed above the matching of the experimental and simulated optical properties of the structure with nanoparticles. The latter shows that having metal nanoparticles on the back helps further enhancing the absorption of a silicon active layer. This confirms that our fabricated nanoparticles are behaving as the simulated nanoparticles and not as a planar back reflector.

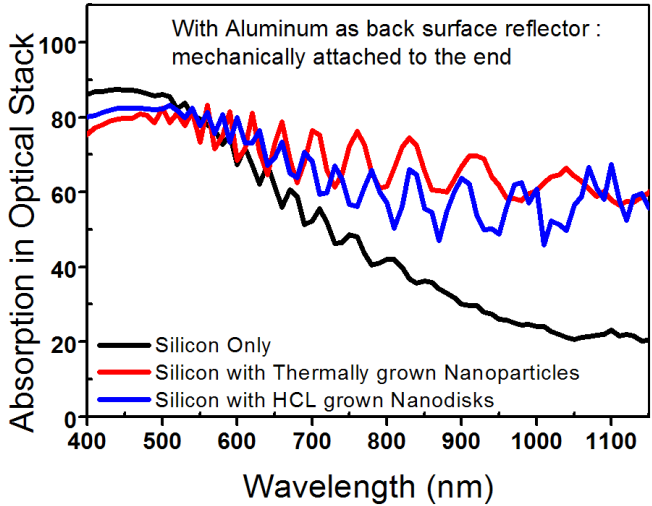

*Figure 13: Absorption in optical stack in Silicon only (black), with random plasmonic nanostructure (red) and with quasi periodic plasmonic nanodisks (blue). In all cases detached aluminium is used as back surface reflector.*

In order to avoid transmission losses we have proposed to add an aluminum layer as back surface reflector on the back of the nanoparticles. Again, absorption measurements show as expected, in Figure 13, that absorption in the structure is highest for the case with random nanoparticles, then for quasi periodic nanoparticles, and minimum for the reference structure.

As we showed in the simulation section above that there is some absorption in the metals, and in particular in the Al mirror, and showed then the potentialities of a back reflector based on dielectric nanoparticles, we now compare the previous structures with a stack having the dielectric nanoparticles coating as back reflector. We obtain a similar result, with a smaller difference between the two kinds of nanoparticles. It is thus likely that part of the supplementary absorption in the case of the annealed random particles was due to enhanced parasitic absorption in the back reflector.

The dielectric nanoparticles coating is known to be a lambertian scatterer. Fig 14 shows the absorption of the stack with a simple $TiO_2$ back reflector and then with the metal nanoparticles sandwiched between the $TiO_2$ and the silicon. The presence of the metal NPs further enhances absorption in silicon, showing that they scatter light beyond the lambertian pattern.

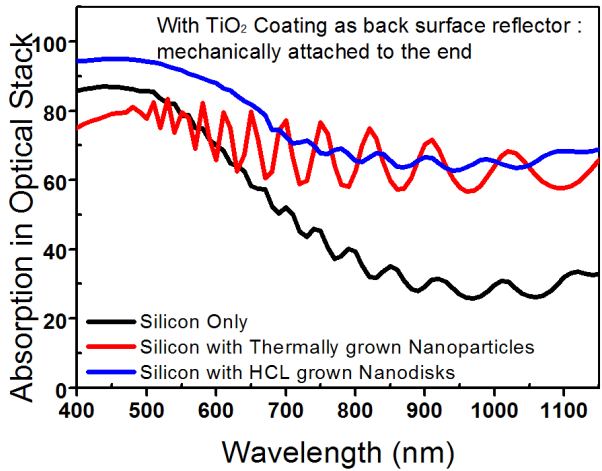

*Figure 14: Absorption in optical stack in Silicon only (black), with random plasmonic nanostructure (red) and with quasi periodic plasmonic nanodisks (blue). In all cases a dielectric nanoparticles coating is used as back surface reflector.*

We finally compare the performances of the various optimized back reflectors: simple metal, nanoparticles with metal back reflector, and metal nanoparticles with a dielectric nanoparticles back reflector. One can see that the overall absorption of the structure is the highest with the last combination (a stack of dielectric and metallic nanoparticles), whereas we have shown in this study that the absorption of the dielectric nanoparticles is negligible. This means that the optimal stack that we deduced in simulation is indeed further improved by the use of a scattering dielectric back reflector.

**Conclusion**

We studied the effect of plasmonic nanoparticles shape, size and embedding materials on their scattering behavior. We then optimized a complete solar cell back reflector scheme by combining plasmonic and dielectric nanoparticles and were able to achieve 69.6% of $J_{scmax}$ enhancement in simulations when comparing to a typical stack with a planar back reflector and a back reflector with plasmonic and dielectric nanoparticles. We can safely conclude based on simulations, and experimental validation of our simulations, that incorporation of nanodisks of optimized parameters in the back reflector scheme, along with a perfect reflector as back surface reflector, results in broadband absorption enhancement in ultra-thin c-Si layer. The optimized structure has a spacer thickness of 10nm, silver nanodisks of height 50nm, diameter 200nm, and capping layer of thickness 210nm. We also show that using a dielectric-nanoparticles coating as back reflector further enhances scattering from the plasmonic back reflector and also avoids any parasitic absorption loss, and should therefore be preferred to a metallic back reflector. We believe that this work paves the way to having an optimal back reflecting scheme for thin c-Si PV cells.

**Acknowledgements**

This work was performed in the frame of the European FP7 project PRIMA grant agreement number 248154. The authors would like to acknowledge fruitful discussions with Sven Leyre (Katholieke Hoge School Ghent).